\documentclass[%
 reprint,
superscriptaddress,
 amsmath,amssymb,
 aps,
prl,
]{revtex4-1}

\usepackage[normalem]{ulem}
\usepackage{graphicx}
\usepackage{dcolumn}
\usepackage{bm}
\usepackage[dvipsnames]{xcolor}

\begin{document}

\title{Image of dynamic local exchange interactions in the dc magnetoresistance of spin-polarized current through a dopant}

\author{Stephen R. McMillan}
\affiliation{Optical Science and Technology Center, and Department of Physics and Astronomy, University of Iowa, Iowa City, Iowa 52242, USA}
\affiliation{Pritzker School of Molecular Engineering, University of Chicago, Chicago, Illinois 60637, USA}
\author{Nicholas J. Harmon}
\altaffiliation{Current address: Department of Physics, University of Evansville, Evansville, Indiana 47722, USA}
\affiliation{Optical Science and Technology Center, and Department of Physics and Astronomy, University of Iowa, Iowa City, Iowa 52242, USA}
\author{Michael E. Flatt\'{e}}
\affiliation{Optical Science and Technology Center, and Department of Physics and Astronomy, University of Iowa, Iowa City, Iowa 52242, USA}
\affiliation{Pritzker School of Molecular Engineering, University of Chicago, Chicago, Illinois 60637, USA}
\affiliation{Department of Applied Physics, Eindhoven University of Technology, P.O. Box 513, 5600 MB, Eindhoven, The Netherlands}

\date{\today}

\begin{abstract}
We predict strong, dynamical effects in the dc magnetoresistance of current flowing from a spin-polarized electrical contact through a magnetic dopant in a nonmagnetic host.  Using the stochastic Liouville formalism we calculate clearly-defined resonances in the dc magnetoresistance when the applied magnetic field matches the exchange interaction with a nearby spin. At these resonances spin precession in the applied magnetic field is  canceled by spin evolution in the exchange field, preserving a dynamic bottleneck for spin transport through the dopant. Similar features emerge when the dopant spin is coupled to nearby nuclei through the hyperfine interaction.  These features provide a precise means of  measuring exchange or hyperfine couplings between localized spins near a surface using spin-polarized scanning tunneling microscopy, without any ac electric or magnetic fields, even when the exchange or hyperfine energy is orders of magnitude smaller than the thermal energy.
\end{abstract}

\maketitle



Localized spin states with long spin coherence times are a fundamental element of the quantum coherent spin systems\cite{Awschalom2002,Hanson2007} that underlie quantum sensing\cite{Degen2017} and quantum information processing\cite{Nielsen2010}. Dopants, and some other controlled defects, in semiconductor or insulator hosts provide a robust realization of these localized coherent spins\cite{Koenraad2011,pla2012}. Optical probes have revealed the spin dynamics of such spin centers, along with their coherent interactions with neighboring spins\cite{Balasubramanian2009,Ohno2012,Dobrovitski2013,Myers2014,Christle2017,Nagy2019}. Similar evidence of all-electrical probing of coherent interactions has been achieved at low temperatures in dopants in large magnetic fields\cite{Broome2018,Pakkiam2018}, and with electron spin resonance combined with spin-polarized scanning tunneling microscopy (STM-ESR)\cite{baumann2015,Bae2018,Willke2018,yang2018}. Low-field magnetoresistance without rf fields in much higher temperature dc electrical measurements, however, has been found in transport from a magnetic contact\cite{Txoperena2014,Inoue2015} through  traps to a nonmagnetic region, as well as in transport through  traps in semiconductor devices\cite{Ashton2019}. These effects can be understood  as originating from coherent spin dynamics in resonant\cite{Song2014}  or non-resonant (incoherent hopping) transport\cite{Inoue2015}, and are largely limited to zero-field peaks or dips in the electrical resistance unless the spin polarization of a contact is time dependent\cite{Harmon2018}.

Here we predict that the {\it dc}  magnetoresistance of  electrical current through the spin state of a single dopant, which can be measured e.g. using spin-polarized STM (SP-STM\cite{wortmann2001,wiesendanger2007,loth2010}), should directly image the coherent effects of the environment on the dopant's spin dynamics, {\it e.g.}  exchange or hyperfine interactions, even at temperatures far larger than the energy scales of those interactions. These sharp environmental features in the dc magnetoresistance will occur in the absence of spin-orbit interaction in the system, and will persist even when the transport process is  charge incoherent (spin orientation is preserved when hopping to and from the dopant's spin state). 
We consider the single dopant to reside in the surface region of a weakly conducting nonmagnetic  host, and under the influence of a second nuclear or electron spin, which affects the first spin through isotropic exchange or isotropic hyperfine interactions. The charge and spin magnitude of the second spin is fixed, as would be appropriate for the electron spin of a deep trap or for a nuclear spin. Strong dc magnetoresistance features are seen when the dopant's spin precession due to the applied magnetic field cancels  the precession due to the exchange or hyperfine interaction with the second spin. These features can be resolved at very small magnetic fields and thus should permit the measurement of milliTesla hyperfine interactions and $\mu$eV exchange interaction strengths, such as are predicted for spin centers in wide-gap semiconductors such as diamond\cite{Kortan2016}. 
Double quantum dot measurements of spin bottlenecks\cite{Petta2005} would not see these features, as the spin polarization {\it transverse} to the applied magnetic field must be measured. 
These coherent features in the magnetoresistance will remain detectable, due to the persistent coherence of the spin, even when the thermal energy exceeds the energy scales of spin interactions with the environment by orders of magnitude, so long as the dopant's spin coherence time exceeds the nanosecond transport times of the junction. This differs from SP-STM measurements of spin excitations during inelastic tunneling spectroscopy\cite{Hirjibehedin2006,Wiesendanger2009}, which cannot be resolved when the temperature exceeds the energy of the spin excitation.
 Similar features appear to occur in the photoluminescence from organic molecules when the applied field brings a triplet exciton state into resonance with the singlet exciton state\cite{Bayliss2018}, although these occur at far larger magnetic fields ($\gg$~Tesla).  More generally, these coherent features may emerge in any flow of current through localized states in nanoscale devices containing contacts with spin polarization oblique to an external field.

\begin{figure}[t]
\begin{centering}
       \includegraphics[width=\columnwidth]{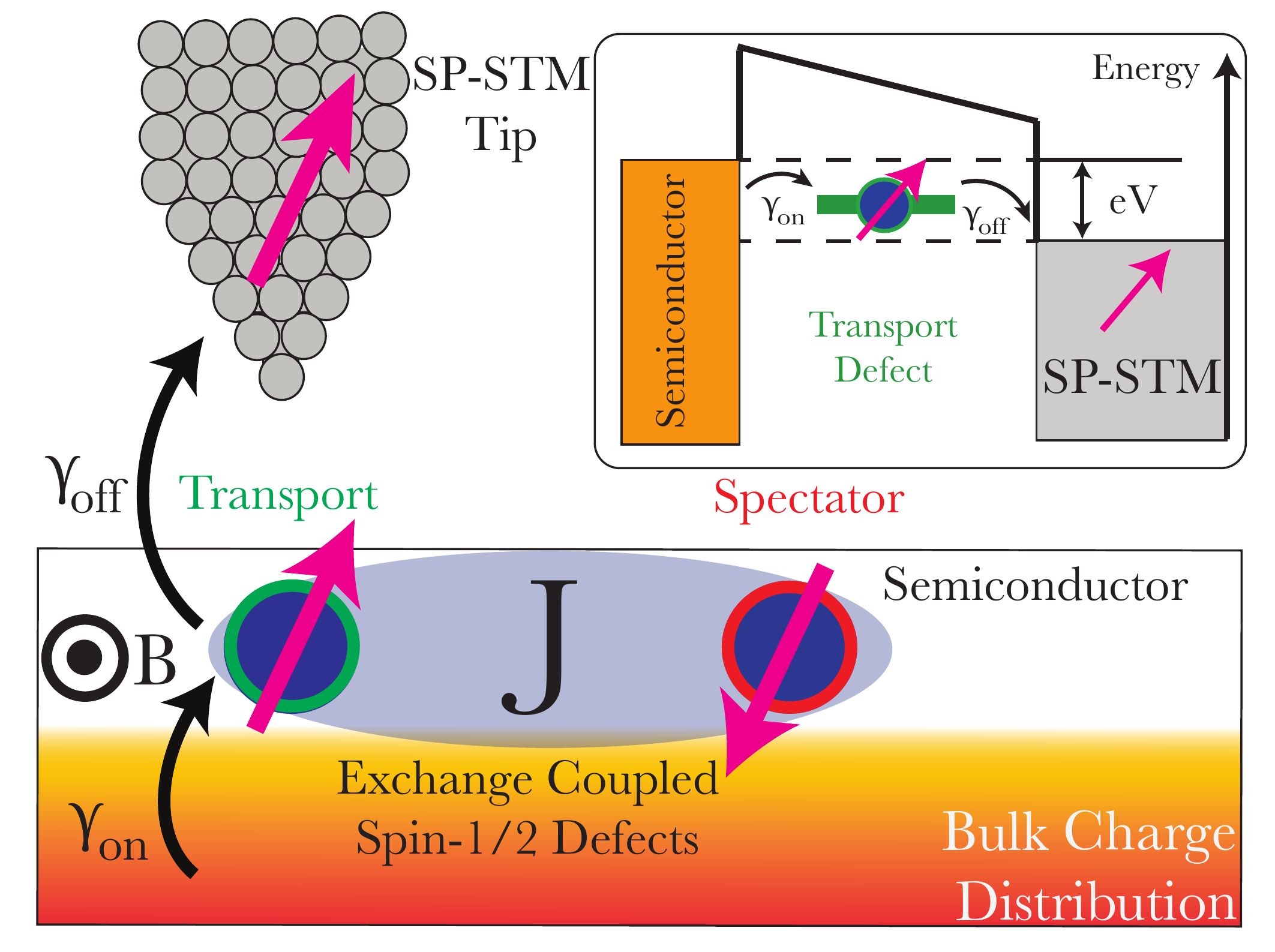}
       \caption[]
{Schematic current path for an electron through a single dopant  (transport site, in green) that is exchange-coupled to another spin-1/2  (spectator site, in red). The transport site has two charge states, empty (spin-0) and full (spin-1/2), and the spectator site's charge state is stable. These spins reside near the surface of the nonmagnetic host and the Coulomb repulsion at the transport site is assumed to be sufficiently large to prevent double occupation (see inset). Hopping from the transport site to a spin polarized scanning tunneling microscope (SP-STM) tip occurs with rate $\gamma_\text{off}$, which is controllable by moving the STM tip relative to the surface. Replenishment of the occupation of the transport site from the nonmagnetic host occurs with a rate $\gamma_\text{on}$. If the host is a semiconductor the charge distribution can be adjusted by the STM voltage $V$ through tip induced band bending, which adjusts $\gamma_\text{on}$ and $\gamma_\text{off}$.}\label{fig:setup}
       \end{centering}
\end{figure}

The detailed configuration is shown in Fig.~\ref{fig:setup}. Here we focus on the concrete situation in which the dopant through which transport occurs (transport site, in green) possesses two states, one of which is spin-0 and ``empty'', and the other is spin-1/2 and ``full'', corresponding to an additional electron (spectator site, in red); similar features are expected for a wide variety of transport and spectator spin states. The charge state of the spectator spin is unaffected during transport. The details of the magnetoresistive curves, and the visibility of coherent features within them, depend on the relative strength of the exchange interaction $J$ and the hopping rates, here $\gamma_{on}$ from the nonmagnetic host to the transport site, and $\gamma_{off}$ from the transport site to the SP-STM tip. The use of the STM provides atomic-scale resolution and permits direct selection of the dopant spin to be probed, and it also provides direct means to independently tune $\gamma_{on}$ and $\gamma_{off}$. By moving the tip towards or away from the surface the $\gamma_{off}$ can be adjusted\cite{zhang2012,tung2001,lee2010}, and through tip-induced band bending both $\gamma_{off}$ and $\gamma_{on}$ can be modified\cite{deRaad2002,Feenstra1987}.  We find the most sensitive configurations correspond to $\gamma_{off}\ll\gamma_{on}$, which is the expected experimental situation under ordinary conditions.
We consider the SP-STM to be polarized along the positive $z$-axis, which can point in any direction relative to the surface normal in Fig.~\ref{fig:setup}.

\begin{figure}[t]
\begin{centering}
       \includegraphics[width=\columnwidth]{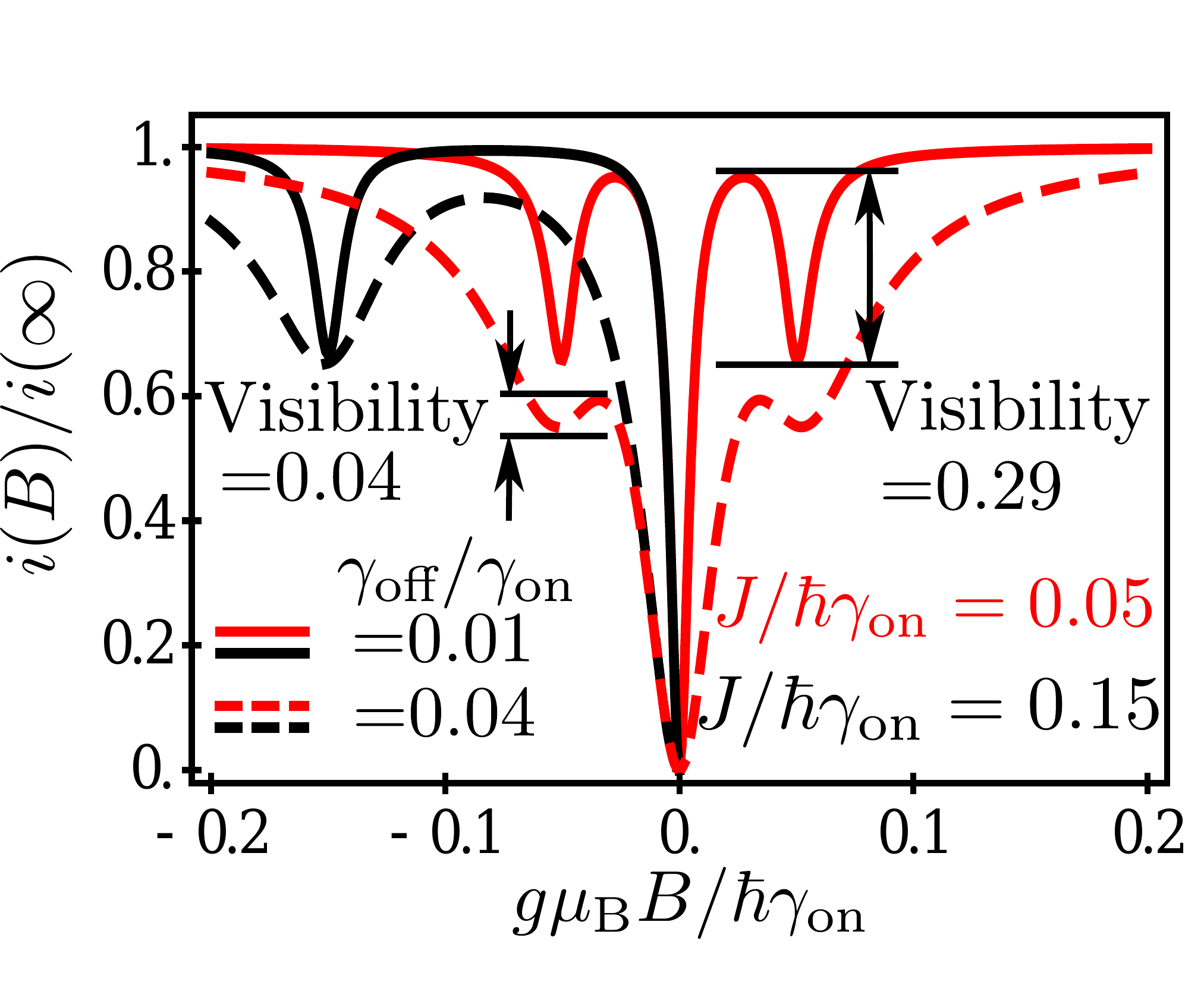}
       \caption[]
{Magnetoresistance of current through spin-1/2 transport dopant for exchange splitting of 0.05 $\hbar \gamma_\text{on}$ (red) and 0.15 $\hbar \gamma_\text{on}$ (black). Features broaden for moderate extraction rates (dashed) compared to slow extraction (solid).}\label{fig:MR}
       \end{centering}
\end{figure}

Our main results are   the predicted finite-field dips (resonances) in the dc magnetoresistance in Fig.~\ref{fig:MR} that occur due to degeneracies in the transport states and the formation of bottlenecks.
Coupling dopants to spin-polarized contacts has been shown to lead to zero-field current bottlenecks resulting from dopant polarization anti-parallel to the magnetization of the contact\cite{Inoue2015}. The application of an external magnetic field, \textbf{B}, transverse to the contact spin polarization, releases the bottleneck by precessing the spin and produces a characteristic dip feature in the zero-field dc magnetoresistance\cite{Inoue2015, Txoperena2014}. Figure~\ref{fig:MR}, however, shows two additional features at finite field when the Zeeman energy of the transport spin equals its exchange energy with the spectator spin. New bottlenecks  emerge under these conditions, because the triplet state polarized parallel or antiparallel to the external field (depending on the sign of $J$) becomes degenerate with the  singlet state ($S$) of these two spins at this external field. We find the dip in current when the Zeeman energy equals $J$ has a maximal value of $1/3$ of the $B=0$ dip. Although our description here is  for the  case of a SP-STM tip that is 100\% spin polarized, these same features will occur for any finite spin polarization, with a reduced amplitude dependent on the tip's spin polarization (as seen for zero-field dips in previous work\cite{Inoue2015}).

The density matrix requires time evolution terms describing coherent evolution of the spins, originating from the magnetic field and exchange, and incoherent processes that inject or extract carriers from the transport site (thus producing current).  We use the stochastic Liouville formalism\cite{kubo1963, vankampen1992}, which has  successfully predicted the manipulation of electroluminescence from organic LEDs\cite{harmon2016b}, exciplex light emission\cite{Wang2016}, spin-pumping\cite{yue2015}, and nuclear spin influences on electron transport\cite{gorman2015}.  The  Hamiltonian  describing coherent evolution of the spins at both the spectator and transport sites in an external magnetic field $\mathbf{B}$ and an exchange interaction $J$ between them is
\begin{equation}
    \hat{H}_{st} = g \mu_\text{B}\mathbf{B}\cdot(\mathbf{S}_s + \mathbf{S}_t) + J \mathbf{S}_s\cdot\mathbf{S}_t, 
\end{equation}
where $\mathbf{S}_{s(t)}$ represents the spin operator at the spectator (transport) site, $g$ is the Land\'e $g$-factor (assumed 2), and $\mu_\text{B}$ is the Bohr magneton. When the transport site is unoccupied the Hamiltonian becomes
\begin{equation}
    \hat{H}_{s} = g \mu_\text{B}\mathbf{B}\cdot\mathbf{S}_s. 
\end{equation}
The coherent evolution of the density operator is
\begin{equation}
    \bigg(\frac{\partial \rho_i(t)}{\partial t}\bigg)_\text{coh} = -\frac{i}{\hbar}[\hat{H}_i,\rho_i(t)],
\end{equation}
where the subscript $i$ is used to denote either the operator for the isolated spectator ($i=s$) or the combined spectator and transport spin ($i = st$) and $[,]$ represents the commutator.

Transport occurs when the dopant at the transport site couples to the unpolarized bulk states and the SP-STM resulting in the generation and extraction of carriers. Processes of this type will not preserve the trace of the individual sub-spaces of the system, but will preserve the total trace, $\text{Tr} \rho_s(t) + \text{Tr} \rho_{st}(t) = 1$, and they will decohere off-diagonal elements of $\rho_{st}$.
The polarization of the SP-STM tip can be written as an operator $\hat{M}_\text{off}(\mathbf{n})$, where $\mathbf{n}$ is the orientation of the SP-STM polarization. In general, for a single spin-1/2 manifold \cite{Farago1971} the interaction operator can be written as a linear superposition of oppositely polarized projection operators $\hat{\pi}(\mathbf{n})$ and $\hat{\pi}(-\mathbf{n})$,
\begin{equation}\label{2x2manifold_int_op_general}
\begin{aligned}
    \hat{m}(\mathbf{n}) &= a \hat{\pi}(\mathbf{n}) + b \hat{\pi}(-\mathbf{n})\\
    &= \frac{1}{2}\{(a+b)\hat{\sigma}_0 + (a-b)\mathbf{n}\cdot\hat{\sigma}\}
\end{aligned}    
\end{equation}
where $a$ and $b$ are interaction amplitudes, and $\mathbf{n}$ is the orientation of the polarization.
In the case of a manifold describing two spin-1/2 dopants the operator results from the direct product of two single spin operators, 
\begin{equation}\label{4x4manifold_int_op}
\begin{aligned}
    \hat{M}(\mathbf{n}) &= \hat{m}_t(\mathbf{n})\otimes\hat{m}_s(\mathbf{n})\\
\end{aligned}    
\end{equation}
where $\hat{m}_t(\mathbf{n}) = a_t  \hat{\pi}(\mathbf{n}) + b_t \hat{\pi}(\mathbf{-n})$ and $\hat{m}_s(\mathbf{n}) = a_s  \hat{\pi}(\mathbf{n}) + b_s \hat{\pi}(\mathbf{-n})$. The coefficients for  $\hat{M}_\text{off}$  are $a^\text{off}_t = 1,\; b^\text{off}_t = 0,\; a^\text{off}_s = b^\text{off}_s = 1/2$. $\hat{M}_\text{on}$  couples the  bulk to the dopant, and for unpolarized bulk states has the coefficients $a^\text{on}_t = a^\text{on}_s = b^\text{on}_t = b^\text{on}_s = 1/2$.
We assume a large on-site Coulomb repulsion that excludes double occupation of the dopant sites. The spin state of the carrier entering the transport site from the bulk is determined by the spin polarization of the bulk ($\hat{M}_\text{on}$) and the spin state of the spectator is determined by the density matrix for the $2\times 2$ manifold, $\rho_s(t)$. These considerations allow one to write down a current operator for the reoccupation of the transport site,
\begin{equation}
    \hat{i}_\text{on} = \frac{e\gamma_\text{on}}{2}\hat{m}^\text{on}_t(\mathbf{n})\otimes \rho_s(t),
\end{equation}
where $e$ is the electronic charge. 

The current from the dopant to the SP-STM depends on the availability of spin states in the tip ($\hat{M}_\text{off}$) and the two-dopant density matrix $\rho_{st}$. The current operator for dopant-to-tip transport is 
\begin{equation}
    \hat{i}_\text{off} =\frac{e\gamma_\text{off}}{2}\{\hat{M}_\text{off},\rho_{st}(t)\},
\end{equation}
where the anticommutator $\{,\}$ guarantees the operator is Hermitian and properly decoheres off-diagonal elements of $\rho_{st}$\cite{Haberkorn1976}. The current onto and off of the transport site is found from the trace of the current operators. The total current onto and off of the transport site must be equal: $\text{Tr}\hat{i}_\text{on} = \text{Tr}\hat{i}_\text{off}=i,$ however  the spin current $i^\text{spin}_\text{on(off)}= \text{Tr}[\hat{i}_\text{on(off)}\mathbf{\sigma}]$ is not conserved due to decoherence.

The appropriate term can now be added to the stochastic Liouville equation for carrier generation 
\begin{equation}
    \bigg(\frac{\partial \rho_{st}(t)}{\partial t}\bigg)_\text{gen} = \gamma_\text{on}\hat{m}^\text{on}_t(\mathbf{n})\otimes \rho_s(t)
\end{equation}
and extraction
\begin{equation}
    \bigg(\frac{\partial \rho_{st}(t)}{\partial t}\bigg)_\text{ext} = -\gamma_\text{off}\{\hat{M}_\text{off},\rho_{st}(t)\}.
\end{equation}
The total equation for the $4\times 4$ density operator $\rho_{st}$ is then 
\begin{equation}\label{SLE4x4}
    \frac{\partial \rho_{st}(t)}{\partial t} = -\frac{i}{\hbar}[\hat{H}_{st},\rho_{st}(t)]+\gamma_\text{on}\mathbf{I}_t\otimes \rho_s(t)-2\gamma_\text{off}\{\hat{M}_\text{off},\rho_{st}(t)\},
\end{equation}
where $\mathbf{I}_t$ is the identity matrix in the transport subspace and the relation $2 \hat{m}^\text{on}_t(\mathbf{n}) = \mathbf{I}_t$ has been used. Eq.~(\ref{SLE4x4}) is coupled to the single spin density operator $\rho_s$ through the generation term. The trace of the total system is preserved and  thus \hbox{$\text{Tr}\rho_s + \text{Tr}\rho_{st} = 1$}; generation(decay) of the $4\times 4$ density matrix must be balanced by the decay(generation) of the $2\times 2$ density matrix allowing one to derive for the isolated spectator,
\begin{equation}
    \frac{\partial \rho_s(t)}{\partial t} = -\frac{i}{\hbar}[\hat{H}_s,\rho_s(t)]+2\gamma_\text{off}\text{Tr}_t\{ \hat{M}_\text{off},\rho_{st}(t)\} -2 \gamma_\text{on}\rho_s(t),
\end{equation}
where $\text{Tr}_t$ is a partial trace over the transport subspace.

These expressions can be solved analytically, but are very cumbersome. A full analytic solution is presented in the Supplemental Material, along with a more intuitive derivation under some approximations. The expressions for the near zero field feature agree with calculations of current through mediating dopants in insulating spacers, during either resonant\cite{Song2014} or incoherent\cite{Inoue2015} transport. However we are interested in the expressions for conditions where the Zeeman energy  ($\tilde{B} = g\mu_B|{\bf B}|$) approaches the exchange energy $J$. An approximate expression valid for $\gamma_\text{on}\gg \tilde{B}, J\gg \gamma_\text{off}$ near the  $\tilde{B}= J$ resonance is 
\begin{equation}
i = \left(\frac{\gamma_\text{off}}{2}\right)\frac{6 J^2 \gamma_{\text{off}}^2+4\left(\tilde{B}^2-J^2\right)^2}{
9J^2 \gamma_{\text{off}}^2+ 4\left(\tilde{B}^2- J^2\right)^2} \label{analyt}
\end{equation}
where the prefactor is the limiting value away from the resonant features. Thus the {\it relative} current has a limiting value for $\tilde{B}=J$ of $2/3$, corresponding to a visibility of $1/3$ as seen below in numerical calculations.
Figure~\ref{fig:MR} shows numerical values for the magnetoresistance, for a 100\%~spin-polarized STM tip. 

\begin{figure}[t]
\begin{centering}
       \includegraphics[width=\columnwidth]{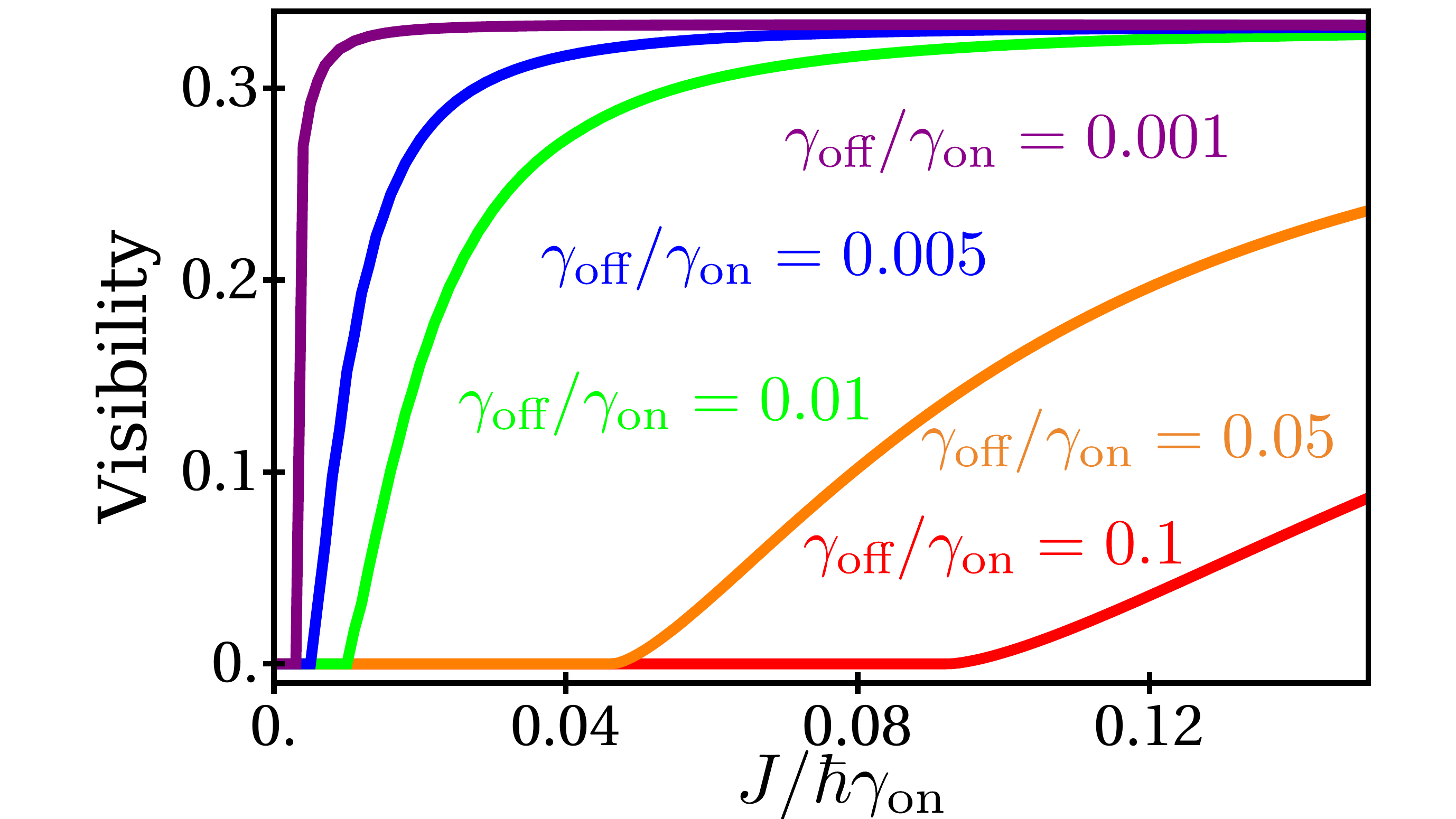}
       \caption[]
{Visibility of the finite-field feature (resonance) as a function of the exchange parameter plotted for different values of the hopping ratio. Smaller hopping ratios yield higher visibility due to extended spin evolution in the exchange field. }\label{fig:visVSexg}
       \end{centering}
\end{figure}

The visibility for different hopping ratios is plotted as a function of the exchange coupling parameter in Fig.~\ref{fig:visVSexg}. At a finite-field dip in current one of the triplet states ($T_+$ or $T_-$ along the direction of \textbf{B}, depending on the sign of $J$) becomes degenerate with the singlet state $S$. The other two triplet states are singly degenerate in the combined Zeeman-exchange Hamiltonian and thus do not participate in a bottleneck configuration. The bottleneck in current manifests as a reduced rate, for one of the two degenerate states, for the transport site's electron to hop to the SP-STM tip. The linear combination of those two states that has the lowest transport-site tunneling probability to the SP-STM tip  is $\big|\phi_b\big>$, and the orthogonal state is $\big|\overline{\phi_b}\big>$, where
\begin{eqnarray}
\big|\phi_b\big> &=& 6^{-1/2}\big|\uparrow \Uparrow\big> + 6^{-1/2}\big|\downarrow \Downarrow\big> + (2/3)^{1/2}\big|\downarrow \Uparrow\big>,\\
\big|\overline{\phi_b}\big> &=& (12)^{-1/2}(\big|\uparrow \Uparrow\big> +\big|\downarrow \Downarrow\big> -\big|\downarrow \Uparrow\big>) + (3/4)^{1/2}\big|\uparrow \Downarrow\big>.\nonumber
\end{eqnarray} 
The first spin (single arrow) represents the transport site and the second (double arrow) represents the spectator site, and the axes are along the SP-STM spin polarization.
$\phi_b$ has a transport spin polarization of $2/3$, which produces a maximum dip in the current of $1/3$ as derived in the Supplemental Material, as plotted in Figs.~\ref{fig:MR} and~\ref{fig:visVSexg}, and as apparent in Eq.~(\ref{analyt}) for $\tilde{B}= J$.

Figure~\ref{fig:visVSexg} also illustrates the increase in resolution and precision  gained by minimizing the hopping ratio. For a dopant-to-tip hopping rate of $\gamma_\text{off} = 1\text{ ns}^{-1}$ we predict exchange energy resolution $\sim \mu$eV. If the energy splittings to double occupancy of the transport site greatly exceed the thermal energy, and the spin coherence times of the spins exceed the timescales set by the smallest energies in the problem (Zeeman energy, exchange energy, or hopping) then the magnetoresistance features survive; observations of zero-field dips are seen at room temperature\cite{Ashton2019}. The energy sensitivity surpasses the practical limits of STM spectroscopic probes of exchange interactions in semiconductor dopant pairs\cite{Flatte2000,Kitchen2006}, and could be measured at temperatures far above where inelastic spin excitations can be imaged for adatoms\cite{Hirjibehedin2006, Wiesendanger2009}.

\begin{figure}[t]
\begin{centering}
       \includegraphics[width=\columnwidth]{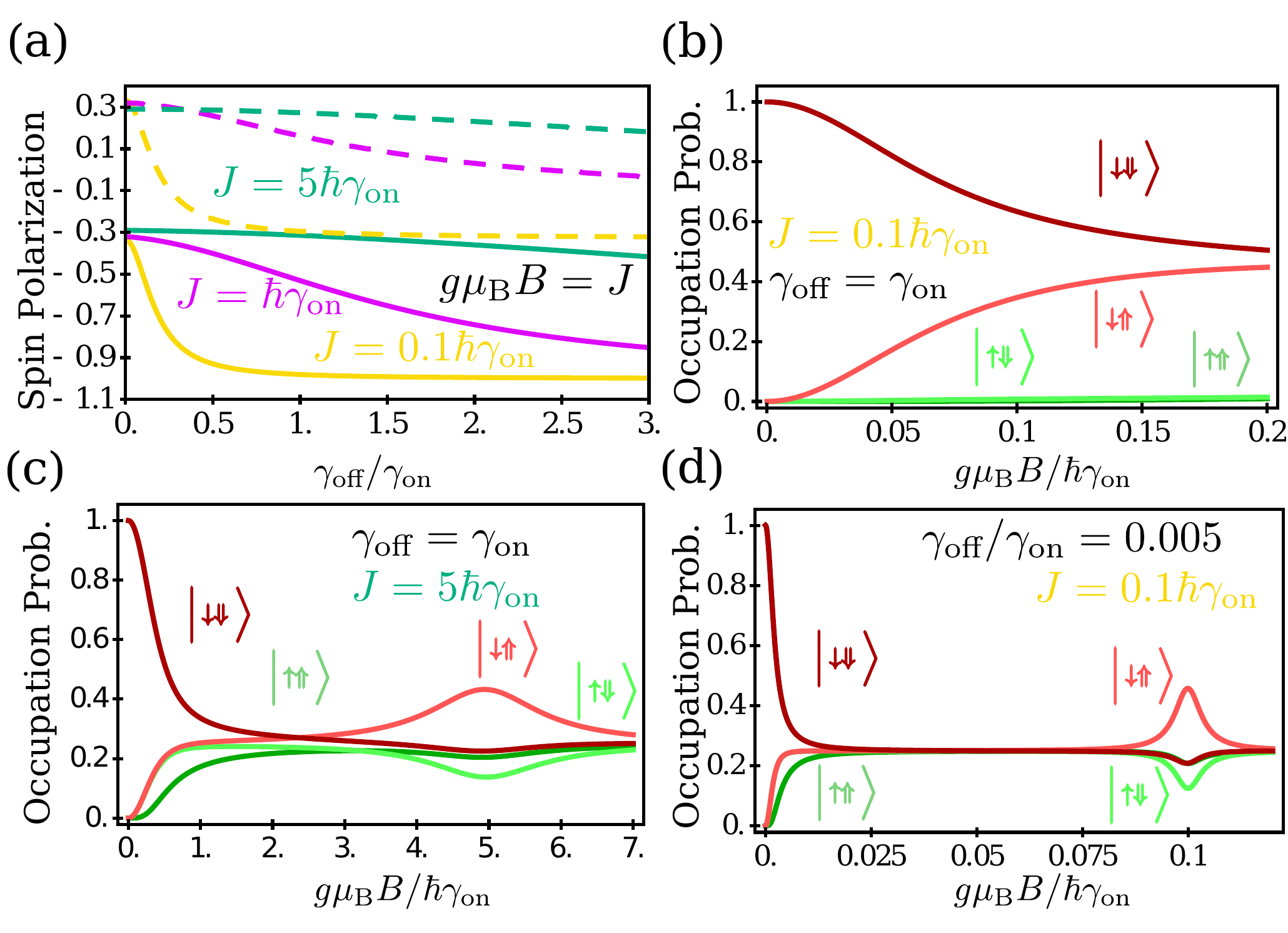}
       \caption[]
{(a) Spin polarization as a function of hopping ratio plotted for different values of the exchange coupling. The dashed lines represent the spectator spin and solid lines represent the transport spin. (b-d) Occupation probability for the different states in the product basis as a function of the transverse magnetic field. Shown are (b) $\gamma_\text{off}=\gamma_\text{on}$,  $\gamma_\text{off}\gg J$, (c) $\gamma_\text{off}=\gamma_\text{on}$, $\gamma_\text{off}\ll J$, and (d) $\gamma_\text{on}\gg J\gg\gamma_\text{off}$. Good visibility is realized when  $\gamma_\text{off}\ll J$, and especially for (d). }\label{fig:spinpolVratio}
       \end{centering}
\end{figure}

Figure~\ref{fig:spinpolVratio}(a) shows calculations for the degree of spin polarization as a function of the hopping ratio for different values of the junction parameters. The dashed lines represent the spectator spin and the solid lines represent the transport spin. In Fig~\ref{fig:spinpolVratio}(b-d) shows the occupation probability for each state in the product basis as a function of $B$. An analytic treatment for $\tilde{B}=J$ and the limit $\gamma_{off}\ll J \ll \gamma_{on}$ [similar to panel (d)], described in the Supplemental Material,  predicts  a  polarization for $\big|\downarrow\Uparrow\big>$ of $11/24$, for $\big|\uparrow\Downarrow\big>$ of $1/8$, and for $\big|\uparrow\Uparrow\big>$ and $\big|\downarrow\Downarrow\big>$ of $5/24$, which are all different from the random average of $1/4$.   These results are very similar to those shown in Fig.~\ref{fig:spinpolVratio}(d). The spectator spin is also dynamically polarized during this process, even when the transport spin state is empty; we find for the limiting conditions above, similar to  Fig.~\ref{fig:spinpolVratio},(d) a polarization of $1/2$ antiparallel to the magnetic field direction and $1/4$ parallel to the SP-STM polarization direction for a total polarization of $\sqrt{5}/4$.

\begin{acknowledgments}
This material is based on work supported by the U. S. Department of Energy, Office of Science, Office of Basic Energy Sciences, under Award Number DE-SC0016447. 
\end{acknowledgments}

\end{document}